\pdfoutput=1
\RequirePackage{ifpdf}
\ifpdf 
\documentclass[pdftex]{sigma}
\else
\documentclass{sigma}
\fi

\numberwithin{equation}{section}

\begin{document}


\renewcommand{\PaperNumber}{012}

\FirstPageHeading

\ShortArticleName{New Variables of Separation for the Steklov--Lyapunov System}

\ArticleName{New Variables of Separation\\ for the Steklov--Lyapunov System}

\Author{Andrey V.~TSIGANOV}

\AuthorNameForHeading{A.V.~Tsiganov}
\Address{St.~Petersburg State University, St.~Petersburg, Russia}
\Email{\href{mailto:andrey.tsiganov@gmail.com}{andrey.tsiganov@gmail.com}}

\ArticleDates{Received October 31, 2011, in f\/inal form March 12, 2012; Published online March 20, 2012}

\Abstract{A  rigid body in an ideal f\/luid  is an important example of Hamiltonian systems on a dual to the semidirect product Lie algebra $e(3) = so(3)\ltimes\mathbb R^3$.  We present the bi-Hamiltonian structure and the corresponding variables of separation on this phase space for the Steklov--Lyapunov system and it's gyrostatic deformation.}

\Keywords{bi-Hamiltonian geometry; variables of separation}

\Classification{70H20; 70H06; 37K10}

\section{Introduction}

Rigid body dynamics in  an ideal incompressible f\/luid  is rich with problems interesting from  mathematical point of view, in particular, the research  of integrable problems. Certainly, the most famous are the three integrable cases under the names of Kirchhof\/f, Clebsch, Steklov and Lyapunov. The latter two cases are more interesting  because there are no a obvious symmetry groups associated with the additional integrals of motion.

All these classical cases were discovered and carefully studied in the 18th and  19th centuries~\cite{bt09}.  For instance, the Kirchhof\/f equations for the Clebsch and  Steklov--Lyapunov cases were f\/irst solved explicitly by K\"otter  after some   mysterious separation of variables~\cite{kot00,kot92}. At the moment no separation which is alternative to his original separation of variables is known for these systems, even though there is a lot of  literature dedicated to the problem, including the theta-functions solutions, associated with the Lax matrices, and the detailed geometric descriptions of the invariant surfaces on which the motions evolve, see books~\cite{avm04, bbe94} and references within.

 In this paper  we apply the  bi-Hamiltonian geometry to the direct calculation of   variables of new separation for the Steklov--Lyapunov system and it's Rubanovsky generalization. We have to  notice right away, that our main purpose is  the development of the bi-Hamiltonian geometry instead of integration of  particular equations of motion using separation of variables method.

An integrable system is separable, if there are  $n$ separation relations
\begin{gather}
\label{seprel}
\Phi_i(u_i,p_{u_i},H_1,\dots,H_n)=0 ,\qquad i=1,\dots,n  ,
\qquad\mbox{with} \quad \det\left[\frac{\partial \Phi_i}{\partial H_j}\right]
\not=0,
\end{gather}
connecting single pairs $(u_i, p_{u_i})$ of canonical variables of separation  with the $n$ functionally independent Hamiltonians $H_1,\ldots, H_n$.  Solving  these  relations  in terms of $p_{u_i}$  one gets the Jacobi equations and the corresponding  additively separable complete integral of the Hamilton--Jacobi equation
\[
W=\sum_{i=1}^n \int^{u_i} p_{u_i}(u'_i, \alpha_1,\ldots,\alpha_n)  du'_i ,\qquad\alpha_j= H_j .
\]
Of course, any variables of separation are determined up to the trivial  transformation
\begin{gather}\label{triv-trans}
u_i\to \tilde{u}_i=f_i(u_i,p_{u_i}),
\end{gather}
which preserve all the properties of algebraic curves def\/ined by separated relations~(\ref{seprel}). Howe\-ver, if we have another separated relations for the same integrals of motion
\[
\Psi_i(v_i,p_{v_i},H_1,\dots,H_n)=0  ,\qquad i=1,\dots,n  ,\qquad\mbox{with} \quad \det\left[\frac{\partial \Psi_i}{\partial H_j}\right]
\not=0  ,
\]
which can not be reduced to initial ones (\ref {seprel}) by trivial change of variables~(\ref{triv-trans}), we usually say about dif\/ferent variables of separation. Of course, any two families of canonical variables on a~given phase space are related by a generic canonical transformations
\begin{gather}\label{gen-trans}
v_i=g_i(u_1,\ldots,u_n,p_1,\ldots,p_n) ,\qquad
p_{v_i}=h_i(u_1,\ldots,u_n,p_1,\ldots,p_n) .
\end{gather}
In contrast with (\ref{triv-trans}) the notion of generic transformations (\ref{gen-trans}) allows us to study  relations between distinct algebraic curves, for instance covering of the algebraic curves (see works of Poincar\'e, Humbert, Frey, Kani, Kuhn, Shaska) or curves with  isogenous Abel varietes (see works of  Richelot, Brock, Hayashida, Nishi, Ibukiyama, Katsura,  van Wamelen). Such relations give us a lot of examples of the reductions of Abelian integrals  (see works of Hermite, Goursat, Burkhardt, Brioschi,  Bolza) and, therefore, they may be a source of new ideas in number theo\-ry, algebraic geometry and modern cryptography  (see works of Tate, Faltings, Zarhin, Lange, McMullen, Merel).

Nevertheless,  motivation for the search of such dif\/ferent variables of separation  do not  pure mathematical, because dif\/ferent variables of separation may be useful in dif\/ferent perturbation theories \cite{ts10k,ts10,ts10b}, as well as in distinct procedures of quantization and various methods of qualitative analysis, etc.

The milestones of the variables separation technique include the works of St\"{a}ckel, Levi-Civita, Eisenhart, Benenti and others. The majority of results was obtained for a very special class of integrable systems, important from the physical point of view, namely for the natural Hamiltonian systems with quadratic in momenta integrals of motion on  cotangent bundles to Riemannian manifolds. The Kowalevski and Chaplygin results on separation of variables for the systems with higher-order integrals of motion  have been missed out of this scheme until recently~\cite{ts10k,ts10,ts10b}.

In the Steklov--Lyapunov case we have quadratic  integrals of motions, but the phase space is
the Poisson manifold  instead of the cotangent bundle to Riemannian manifold.  So,  in this case we can  use neither the Levi-Civita criteria, nor the Eisenhart--Benenti theory. Below we show how  variables of separation for the given integrable system may be calculated without any additional information (Killing tensors, Lax matrices, $r$-matrices, links with soliton equations etc.).

 \section[Steklov-Lyapunov system]{Steklov--Lyapunov system}

 Following Kirchhof\/f, we consider the potential motion of a f\/inite rigid body  submerged in an inf\/initely large volume of irrotational, incompressible, inviscid f\/luid that is at rest at inf\/inity, so that  the induced motion of particles
of the f\/luid is completely determined by the motion of the body~\cite{kirh74}.
In this case,  the motion of  rigid body  is described by the classical Kirchhof\/f equations
\begin{gather}\label{Kirh-Eq}
 \dot{M}=M \times
\Omega+p \times U, \qquad \dot{p}=p \times \Omega,
\end{gather}
here $x\times y$ stands for the vector product of three-dimensional vectors.
Vectors $M $ and $p$ are the impulsive momentum and the impulsive force while~$\Omega$ and~$U$ are the angular and linear velocities of the body. All these vectors in~${\mathbb R}^3$ are expressed in the body frame attached to the body  originating at the center of buoyancy~\cite{kirh74}.

 A rigid body in the ideal f\/luid is an  important example of Hamiltonian systems on a dual to  Lie algebra  $e(3)=so(3)\ltimes \mathbb R^3$.  The dual  space $e^*(3)$ is  the Poisson manifold  endowed with  the  canonical  Lie--Poisson brackets
\begin{gather}\label{e3}
 \bigl\{M_i ,M_j \bigr\}=\varepsilon_{ijk}M_k , \qquad
\bigl\{M_i ,p_j \bigr\}=\varepsilon_{ijk}p_k  , \qquad
\bigl\{p_i ,p_j \bigr\}=0 ,
\end{gather}
where $\varepsilon_{ijk}$ is a totally skew-symmetric tensor. There  are two Casimir elements
 \begin{gather}\label{Caz}
C_1=\langle p,p\rangle=|p|^2 \equiv\sum_{i=1}^3p_i^2, \qquad C_2=\langle p,M \rangle\equiv\sum_{i=1}^3p_iM_i ,
\end{gather}
where $\langle x,y\rangle$ means scalar product of two three-dimensional vectors $x,y\in \mathbb R^3$.

As usual  \cite{kirh74},  element   $M\in so(3)$ is identif\/ied with three-dimensional vector $M\in\mathbb R^3$  using  well  known isomorphism of the Lie algebras $(\mathbb R^3, \times)$ and $so(3)$, $[\cdot,\cdot]$
\begin{gather}\label{trans-M}
 z=\left(z_1,z_2,z_3\right)\to {z}_{\mu}= \begin{pmatrix}
            0 & z_3 & -z_2 \\
            -z_3 & 0 & z_1 \\
            z_2 & -z_1 & 0
          \end{pmatrix},
\end{gather}
where $\times$ is a cross product,  $[\cdot,\cdot ]$ is a matrix commutator and index ${\mu}$ means a $3\times 3$ antisymmetric matrix associated with the vector $z$. Using this agreement we can rewrite  canonical Poisson bivector  on $e^*(3)$  in the following compact form
\begin{gather} \label{p-e3}
P=\begin{pmatrix}
      0 & 0 & 0 & 0 & p_3 & -p_2 \\
      0 & 0 & 0 & -p_3 & 0 & p_1 \\
      0 & 0 & 0 & p_2 & -p_1 & 0 \\
      0 & p_3 & -p_2 & 0 & M_3 & -M_2 \\
      -p_3 &0 & p_1 & -M_3 & 0 & M_1 \\
      p_2 & -p_1 & 0 & M_1 & -M_1 & 0
   \end{pmatrix}
   =\begin{pmatrix}
              0 & p_{\mu} \\
              p_{\mu} & M_{\mu}
            \end{pmatrix}
.
\end{gather}
The Hamilton function $H = H (p,M)$ and Lie--Poisson brackets (\ref{e3}) allow us to def\/ine  the Hamiltonian equations of motion
\begin{gather}\label{Kirh-Eqh}
\dot{M}=M \times \dfrac{\partial H}{\partial M}+p \times \dfrac{\partial  H}{\partial p},  \qquad \dot{p}=p
\times \dfrac{\partial H}{\partial M}.
\end{gather}
These generic Euler's equations on $e^*(3)$ coincide with the Kirchhof\/f equations (\ref{Kirh-Eq}),
if $H(p,M)$ is a second-order polynomial  in variables $M$ and $p$.

\begin{remark}
The Lie--Poisson dynamics on $e^*(3)$ can be interpreted as resulting from
reduction by the symmetry group $E(3)$ of the full dynamics on the
twelve-dimensional phase space $T^*E(3)$. Here, symmetry means that the
Hamiltonian that describes the dynamics in $T^*E(3)$ is an invariant to actions
of $E(3)$, i.e., one can translate the inertial frame or rotate it in any direction
without af\/fecting the equations of motion~\cite{nov81}.
\end{remark}

The Steklov--Lyapunov case of the rigid body motion  is characterized by the
following second-order homogeneous polynomial integrals of motion
\begin{gather}
 H_1 = \langle M,M\rangle-2\langle \mathbf
A p,M\rangle-  \langle(\mathbf A^2+2
\mathbf A^{\vee})p,p\rangle  ,
\nonumber  \\
\label{int-stl}
 H_2 = \langle\mathbf A M,M\rangle+2\langle\mathbf A^\vee
p,M\rangle-\bigl( \langle\mathbf A^3p,p
\rangle-\mbox{tr}\,\mathbf A^2\langle\mathbf A p,p \rangle
\bigr) ,
\end{gather}
where wedge denotes an adjoint matrix, i.e.\ a cofactor matrix $\mathbf A^{\vee}=({\det \mathbf A}) \mathbf
A^{-1}$. For integrability $\mathbf A$ has to be  symmetric matrix, which may be reduced to the diagonal form
\begin{gather}\label{a-mat}
\mathbf A=\mbox{diag}(a_1,a_2,a_3)  ,\qquad a_i\in\mathbb R ,
\end{gather}
using  linear canonical transformations of $e^*(3)$. From physical point of view it means that the body axes can always be chosen so that $\mathbf A$ is diagonal.
\begin{remark}
In  \cite{stek93} Steklov found  integrable Hamiltonian $H_2$, whereas  Lyapunov  proved integrability  of the Kirchhof\/f equations with Hamiltonian~$H_1$ in~\cite{lyap97}.  The generic family of  the Steklov--Lyapunov integrable systems was studied by Kolosov in~\cite{kol19}.
\end{remark}

\subsection{Separation of variables by  K\"otter}

The explicit integration of the classical Steklov--Lyapunov systems
via separation of variables had been f\/irst made by F.~K\"otter in 1900~\cite{kot00}.
Here we want to add some new  details in  the known coincidence of the  K\"otter variables of separation~$v_{1,2}$ with the elliptic coordinates on the sphere.

According to \cite{ts05,ts04} there is a Poisson map, which identif\/ies the Steklov--Lyapunov system with  a system that describes motion on the surface of a unit two-dimensional sphere  $\mathbb S^2$  in a~fourth-degree polynomial potential f\/ield. This dynamical system  is separable in  standard elliptic coordinates  on the sphere, and  the inverse Poisson map allows us to get complete solution of the Steklov--Lyapunov system.
\begin{proposition}\label{proposition1}
 If  $\mathbf B=\mathrm{tr}\,\mathbf A-\mathbf A$ and $\mathbf C=\sqrt{\mathbf B^2-4\mathbf A^\vee}$ are  diagonal  matrices with entries
\begin{gather*}
\mathbf B=
                                           \begin{pmatrix}
                                             a_2+a_3 & 0 & 0 \\
                                             0 & a_1+a_3 & 0 \\
                                             0 & 0 & a_1+a_2
                                           \end{pmatrix},\qquad
\mathbf C=
                                           \begin{pmatrix}
                                             a_2-a_3 & 0 & 0 \\
                                             0 & a_3-a_1 & 0 \\
                                             0 & 0 & a_1-a_2
                                           \end{pmatrix},
\end{gather*}
then  the Poisson map $f:(p,M)\to (x,J)$, defined by
\begin{gather}\label{fg}
x=\dfrac{(M-\mathbf B p)\times  p}{|(M-\mathbf B p)\times
p |},\qquad
  J=M+\mathbf C \bigl[  x,x\times  p \bigr]_+ ,
\end{gather}
where
\[{[y,z]_+}_i=\sum_{j,k=1}^{n=3}|\varepsilon_{ijk} |  y_j z_k ,
\]
relates  manifold $e^*(3)$ with coordinates $(p,M)$ and cotangent bundle  $T^*\mathbb S^2$ to the unit two-dimensional sphere $\mathbb S^2$ with coordinates $(x,J)$.
\end{proposition}

The proof consists in the verif\/ication of the Lie--Poisson brackets between variables  $x$ and $J$
\[
 \bigl\{J_i ,J_j \bigr\}=\varepsilon_{ijk}J_k , \qquad
\bigl\{J_i ,x_j \bigr\}=\varepsilon_{ijk}x_k  , \qquad
\bigl\{x_i ,x_j \bigr\}=0 ,
\]
and calculation of  the corresponding   Casimir functions
\[
\langle x,x\rangle=|x|^2=1,\qquad\mbox{and}\qquad \langle x,J\rangle=0 .
\]
So, this coajoint orbit  of $e^*(3)$ with coordinates  $x$ and $J$ is simplectomorphic  to  cotangent bundle~$T^*\mathbb S^2$ to the unit two-dimensional sphere $\mathbb S^2$, see for instance~\cite{nov81}.

Inverse Poisson map  $f^{-1}:(x,J)\to(p,M)$ looks like
\begin{gather}\label{inv-fg}
p=\alpha J+\beta (x\times J) ,\qquad M= J-\mathbf C \bigl[  x,x\times  p \bigr]_+,
\end{gather}
where functions $\alpha$, $\beta$ on $x$, $J$ are solutions of the following  equations
\begin{gather*}
C_1=\langle p,p \rangle = \alpha^2|J|^2+\beta^2|x\times J|^2 ,
\\
C_2=\langle p,M\rangle = \alpha^2\bigl(|J|^2 \langle x,\mathbf A x\rangle-\langle (x\times J), \mathbf A (x\times J)\rangle\bigr)+2\alpha\beta
\langle J, \mathbf A (x\times J)\rangle \\
\phantom{C_2=\langle p,M\rangle =}{}
 + \beta^2\bigl(
\langle (x\times J), \mathbf B (x\times J)\rangle-2\langle J,\mathbf A J\rangle\bigr)+\alpha |J|^2 .
\end{gather*}

\begin{proposition}
The Poisson map  \eqref{fg}  relates  the Steklov integral of motion  $H_1(p,M)$ \eqref{int-stl}  with
the  natural Hamilton function on $T^*\mathbb S^2$
\begin{gather}
H_1(x,J) = \langle J,J \rangle +4\langle x,\mathbf B x\rangle\bigl((\mathrm{tr}\,\mathbf A C_1-C_2)
-C_1\langle x,\mathbf B x\rangle\Bigr)+4C_1\langle x,\mathbf A^\vee x\rangle\nonumber\\
\label{H-S}
\phantom{H_1(x,J) =}{} + 2\,\mathrm{tr}\,\mathbf A C_2-\mathrm{tr}(\mathbf A^2+4\mathbf A^\vee) C_1 .
\end{gather}
The Lyapunov integral  $H_2(p,M)$ \eqref{int-stl} is equal to
\begin{gather}
{H}_2(x,J)=\langle J,\mathbf A J\rangle -4\langle
x,\mathbf A^\vee x \rangle \bigl(C_1\langle x,\mathbf B x
\rangle-(\mathrm{tr}\, \mathbf A C_1-C_2)\bigr)\nonumber\\
\phantom{{H}_2(x,J)=}{}
+2\, \mathrm{tr}\,\mathbf A C_2-2\det\mathbf A C_1 .\label{H2-S}
\end{gather}
Here $C_{1,2}$ are values of the Casimir functions \eqref{Caz}.
\end{proposition}

 The proof of this proposition and all the details may be found in~\cite{ts05,ts04}.

\begin{remark}
It is well-known that  the generic level sets of the Casimir functions (coadjoint orbits) on $e^*(3)$ are only dif\/feomorphic to the cotangent bundle $T^*\mathbb S^2$. It allows us to directly connect  the Kirchhof\/f equations with  the equations of motion by geodesic on  $\mathbb S^2$,  however, at $C_2\neq 0$
we have to  destroy the standard symplectic structure on $T^*\mathbb S^2$ by adding some ``monopole'' terms~\cite{nov81}.

In the Steklov--Lyapunov case we  use the Poisson map~(\ref{fg}), which preserves the standard symplectic structure on  $T^*\mathbb S^2$. As a punishment for this preservation of the  standard symplectic structure we have to consider potential motion on~$\mathbb S^2$ instead of the geodesic motion.
\end{remark}

The Hamiltonians $H_{1,2}(x,J)$ (\ref{H-S}), (\ref{H2-S})  on $T^*\mathbb S^2$ are separable in the elliptic (spheroconical) coordinates ${v}_{1,2}$, which are zeroes of the function
\begin{gather}\label{ell-S}
e(\lambda)=\dfrac{(\lambda-{v}_1)(\lambda-{v}_2)}{\det(\lambda
-\mathbf A)} =\langle x, (\lambda  -\mathbf A)^{-1}x \rangle\equiv\sum_{i=1}^3\dfrac{x_i^2}{\lambda-a_i} .
\end{gather}
This variables of separation satisfy the following separated relations
\begin{gather}\label{sepeq-K}
\Phi(v_i,p_{v_i})=\det(v_i-\mathbf A) p_{v_i}^2+\langle \ell(v_i),\ell(v_i)\rangle=0 ,\qquad i=1,2,
\end{gather}
where the three-dimensional vector  $\ell(\lambda)$ is the so-called K\"otter vector with entries
\begin{gather*}
\ell_i(\lambda)=\dfrac{\sqrt{\lambda-a_i}}{2} \bigl(M_i+\bigl(2\lambda+a_i-\mbox{tr}\,\mathbf A\bigr)p_i\bigr)
\end{gather*}
depending on the auxiliary variable $\lambda$ (spectral parameter). The explicit solution of the corresponding Abel--Jacobi equations  in theta-functions  was given in~\cite{kot00}.  In order get  initial variables~$p$,~$M$ as functions on time variable we have to substitute solutions of the Abel--Jacobi equations into the  variables on $T^*\mathbb S^2$
\begin{gather*}
x_i=\sqrt{\dfrac{(v_1-a_i)(v_2-a_i)}{(a_j-a_i)(a_k-a_i)}},\qquad
J_i=\dfrac{2\varepsilon_{ijk}x_jx_k(a_j-a_k)}{v_1-v_2}\bigl((a_i-v_1)p_{v_1}-(a_i-v_2)p_{v_2}\bigr) ,
\end{gather*}
where $(i,j,k)$ is  permutation of $ (1, 2,3)$, and then into the variables~$p$ and~$M$~(\ref{inv-fg}). Another modern verif\/ication of the K\"otters calculations may be found in~\cite{bf03,fed09}.

\section{Calculation of the variables of separation\\ in bi-Hamiltonian geometry}

We can only guess  how K\"otter  invented the variables of separation $v_{1,2}$, which coincide with  the elliptic coordinates on the auxiliary two-dimensional sphere, because he gave no  explanations of calculations in very brief communication~\cite{kot00}.  It is clear that behind the striking formulas there must be a certain geometric idea, but the domain of applicability of this idea is usually restricted by a partial model under consideration. For instance, we can not apply the K\"otter separation  to the  Kowalevski  top  and  Kowalevski separation to the Steklov--Lyapunov system etc.

Our aim  is to discuss some algorithm of  calculation of the variables of separation in the framework of the bi-Hamiltonian geometry, which is applicable  to many known integrable systems \cite{fp02,ts10k,ts08a,ts08,ts10,ts10b,ts07}. In fact this algorithm consists of the following steps:
\begin{itemize}\itemsep=0pt
 \item calculate the second Poisson bracket compatible with canonical one starting with the given integrals of motion in the involution with respect to this canonical Poisson bracket;
 \item if the Poisson brackets have dif\/ferent symplectic leaves, calculate a projection of the second bracket on symplectic leaves of the f\/irst bracket;
 \item calculate coordinates of separation as eigenvalues of the corresponding recursion operator;
 \item calculate the canonically conjugated momenta with respect to the  f\/irst Poisson bracket;
 \item calculate the separated relations.
\end{itemize}
The input of algorithm  is a set of integrals of motion and canonical Poisson bracket, whereas output is a set of separated relations. 
All details about construction of a suitable projection are discussed in~\cite{fp02}.

Because variables of separation (\ref{seprel}) are def\/ined up to canonical transformations $ u_i\to f(u_i,p_{u_i})$ on the f\/irst step of this algorithm we have to  narrow the search space using some artif\/icial tricks. It is a main technical problem of this method. The second technical problem is  the calculation of the  momenta
conjugated to obtained coordinates, see~\cite{fp02,ts10k,ts10,ts10b}.

\subsection[Polynomial and rational Poisson brackets on $e^*(3)$]{Polynomial and rational Poisson brackets on $\boldsymbol{e^*(3)}$}

Bi-Hamiltonian structures can be seen as a dual formulation of integrability and separability, in the sense that they substitute a hierarchy of compatible Poisson structures to the hierarchy of functions in involution, which may be treated either as integrals of motion or as variables of separation. So, our f\/irst step  is calculation of the second Poisson bivector $P'$ compatible with kinematic Poisson bivector $P$.

According to \cite{ts08a,ts07} any  separable system is  a bi-integrable system, i.e.\   integrals of motion~$H_{k}$~(\ref{seprel}) are  in  bi-involution
\begin{gather}\label{bi-inv}
\{H_i,H_k\}=\{H_i,H_k\}'=0v,\qquad i,k=1,\ldots,n,
\end{gather}
with respect to  compatible  Poisson brackets $\{\cdot,\cdot \}$
and   $\{\cdot,\cdot \}'$ associated with the Poisson bivectors~$P$ and~$P'$, so that
\begin{gather}\label{comp-eq}
[\![P,P]\!]=0,\qquad [\![P,P']\!]=0,\qquad [\![P',P']\!]=0.
\end{gather}
Here $[\![\cdot,\cdot ]\!]$ is the  Schouten bracket. The def\/inition of the  second bracket $\{\cdot,\cdot \}'$  in term of variables of separation may be found in~\cite{ts08a, ts07}.

For the given  integrable system f\/ixed by a kinematic bivector $P$ and  a tuple of integrals of motion $H_1,\ldots,H_n$    bi-Hamiltonian  construction of  variables of separation  consists in a direct solution of the equations (\ref{bi-inv}) and (\ref{comp-eq})  with respect to an unknown  bivector $P'$. The main problem is that the geometrically invariant equations  (\ref{bi-inv}),~(\ref{comp-eq}) have a'priory  inf\/inite number  of solutions \cite{ts10k,ts08a, ts08,ts10,ts10b, ts07}.

In order to get a search algorithm of  ef\/fectively computable  solutions we have to narrow the search space by using some non-invariant additional assumptions. According to \cite{ts08a,ts08}  hereafter we assume that   $P'$ has the same foliations by symplectic leaves as  $P$, i.e.\ that
\begin{gather}\label{add-assum}
P'dC_{1,2}=0
\end{gather}
and $P'$ doesn't have any other Casimir elements. The geometric meaning of this restriction is discussed in~\cite{ts08a, ts08}. In fact it allows us to avoid calculations of the  projection of the second bracket on  the symplectic leaves of the f\/irst bracket.

In the Steklov--Lyapunov case solving equations (\ref{bi-inv}), (\ref{comp-eq}) and (\ref{add-assum}) in the space of homogeneous second-order polynomial bivectors  and  of rational bivectors with the second-order homogeneous  numerators and linear denominators we obtain the following two propositions.

\begin{proposition}\label{proposition3}
If  $c$ and $d$ are two numeric three-dimensional vectors, so that $\langle c,c\rangle=0$, then
equations \eqref{comp-eq} and \eqref{add-assum} on $e^*(3)$ have  a
polynomial  solution
\begin{gather}\label{p2-p}
P'_1= \begin{pmatrix}
\langle c,p\rangle p_{\mu} &
\langle c,M\rangle p_{\mu}+ (p\times M)\otimes c+\frac{1}{2}\left(\frac{1}{\alpha}+\alpha\langle c,d\rangle\right)
\bigl(p\otimes p -\langle p,p\rangle\bigr)\\ & {} +\alpha(c\times p)\otimes (d\times p)
\vspace{1mm}\\
*
& \langle c,M\rangle M_{\mu}+\langle d,p \rangle p_{\mu}
+\frac{1}{2}\left(\frac{1}{\alpha}+\alpha\langle c,d\rangle \right)(p\times M)_{\mu}\\
& {} -\alpha\bigl((c\times p)\times(d\times M)\bigr)_{\mu}
\end{pmatrix} ,
\end{gather}
$\alpha\in\mathbb C$, and a rational solution
\begin{gather}\label{p2-r}
P'_2=\dfrac{1}{\langle c,p\rangle} P'_1+\dfrac{1}{\langle c,p\rangle} \left(\langle c,M\rangle+\dfrac{\alpha(1+\alpha^2\langle c,d\rangle)}{2\alpha^2\langle c,d\rangle} \langle c\times p,d\rangle\right) P ,
\end{gather}
compatible to each other, i.e.\  $[\![P'_1,P'_2]\!]=0$.
\end{proposition}

As above  $\times$ is a cross product,   the antisymmetric matrix $z_\mu$ is def\/ined by  vector $z$  (\ref{trans-M}) and  the  matrix $(x\otimes y)_{ij}=x_iy_j$ is determined by a pair of vectors $x$ and $y$.
In order to explain this notations we write out the corresponding Poisson  brackets
\begin{gather}
\{p_i,p_j\}'_1 = \varepsilon_{ijk}\langle c,p\rangle p_k ,
\nonumber\\
\{p_i,M_j\}'_1 = \varepsilon_{ijk}\langle c,M\rangle p_k+(p\times M)_ic_j+\frac{1}{2\alpha}\bigl(1+\alpha^2\langle c,d\rangle \bigr)\left (p_ip_j-\sum_{l=1}^3 p_l^2\right)\nonumber\\
\phantom{\{p_i,M_j\}'_1 =}{}
+\alpha(c\times p)_i (d\times p)_j,
\label{poi-br1}\\
\{M_i,M_j\}'_1 =\varepsilon_{ijk}\bigl(\langle c,M\rangle  M_k+\langle d,p\rangle  p_k\bigr)\nonumber\\
\phantom{\{M_i,M_j\}'_1 =}{}
 + \varepsilon_{ijk}\left(\frac{1}{2\alpha}\left(1+\alpha^2\langle c,d\rangle \right)(p\times M)_k
-\alpha\bigl((c\times p)\times(d\times M)\bigr)_k\right) .\nonumber
\end{gather}
The second brackets are equal to{\samepage
\begin{gather}
\{p_i,p_j\}'_2 = \varepsilon_{ijk} p_k ,\nonumber\\
\{p_i,M_j\}'_2 = \dfrac{\{p_i,M_j\}'_1}{\langle c,p\rangle}+\dfrac{\varepsilon_{ijk} p_k}{\langle c,p\rangle} \left(\langle c,M\rangle+\dfrac{\alpha(1+\alpha^2\langle c,d\rangle)}{2\alpha^2\langle c,d\rangle} \langle c\times p,d\rangle\right) ,
\label{poi-br2}\\
\{M_i,M_j\}'_2 = \dfrac{\{M_i,M_j\}'_1}{\langle c,p\rangle}+\dfrac{\varepsilon_{ijk} M_k}{\langle c,p\rangle} \left(\langle c,M\rangle+\dfrac{\alpha(1+\alpha^2\langle c,d\rangle)}{2\alpha^2\langle c,d\rangle} \langle c\times p,d\rangle\right) .\nonumber
\end{gather}
We have to stress that this brackets are def\/ined over complex f\/ield because $\langle c,c \rangle=c_1^2+c_2^2+c_3^2=0$.}

Substituting the Poisson brackets $\{\cdot,\cdot \}'_{1,2}$ (\ref{poi-br1}), (\ref{poi-br2}) into~(\ref{bi-inv}) and  solving the resulting equations in the space of the second-order homogeneous polynomials $H_{1,2}$ one gets the following proposition.

\begin{proposition}
The Steklov--Lyapunov integrals of motion $H_{1,2}(p,M)$ \eqref{int-stl} satisfy  the equation~\eqref{bi-inv} at  $\alpha=1$ and
\begin{gather}
c = \dfrac{-1}{ \sqrt{(a_1-a_2)(a_2-a_3)(a_3-a_1)}}\left( \sqrt{a_2-a_3}, \sqrt{a_3-a_1}, \sqrt{a_1-a_2}\right) ,\nonumber\\
\label{ab-v}
d =  \sqrt{(a_1-a_2)(a_2-a_3)(a_3-a_1)}\left( \dfrac{1}{\sqrt{a_2-a_3}}, \dfrac{1}{\sqrt{a_3-a_1}}, \dfrac{1}{\sqrt{a_1-a_2}}\right).
 \end{gather}
 \end{proposition}

 The proof is a straightforward calculation.
\begin{remark}
In fact, polynomial bivector $P'_1$ has been obtained in~\cite{ts08} as an incidental result by investigation of the Poisson bivectors on the Lie algebra $so^*(4)$ and the corresponding integrable cases in the Euler equations on~$so^*(4)$. Now we recover this bivector by solving equations~(\ref{bi-inv}),~(\ref{comp-eq}) and~(\ref{add-assum}) for the Steklov--Lyapunov system.
\end{remark}

Let us brief\/ly discuss the bi-Hamiltonian structure related with the K\"otter variables of separation.
 Elliptic coordinates on the sphere $\mathbb S^2$ (\ref{ell-S}) are associated with the polynomial Poisson bivector
 \[
P'_{ell}=\mathcal L_{X}P,
\]
where $\mathcal L_{X}$ is a Lie derivative along the vector f\/ield  $X=\sum X^j\partial_j$ with the following entries:
\[
X^i=0,\qquad X^{i+3}=\bigl[x\times\mathbf A (x\times J)\bigr]_i,\qquad i=1,2,3 .
\]
Bivector $P'_e$ is compatible with $P$ and has the same foliation by symplectic leaves as~$P$.

Using the Poisson map (\ref{fg}), (\ref{inv-fg}) we can easily express $P'_{ell}$ in the initial  variables~$p$,~$M$. It will be a rational bivector $P'_{ell}=R/Q$, where $R$ is a bivector with fourth-order homogeneous polynomial entries and  $Q=|(M-\mathbf B p)\times p|^2$ is a fourth-order polynomial as well. So, we could directly calculate the K\"otter variables solving equations~(\ref{bi-inv}), (\ref{comp-eq}) and~(\ref{add-assum}) in the corresponding space of rational bivectors.

\subsection{Calculation of variables of separation}

The bi-involutivity of the integrals of motion (\ref{bi-inv})
is equivalent to the existence of  control matrix~$F$ def\/ined by
\begin{gather*}
P'{{dH}}=P\bigl(F{{dH}}\bigr),\qquad\mbox{or}\qquad
P'dH_i=P\sum_{j=1}^n F_{ij} dH_j,\qquad i=1,\ldots,n.
\end{gather*}
The additional assumption  (\ref{add-assum}) ensures that $F$ is a non-degenerate matrix and  the eigenvalues of~$F$ are the desired variables of separation \cite{ts08a, ts08}. Moreover, for the so-called St\"ackel sepa\-rable systems the suitable normalized left eigenvectors of the  control matrix~$F$ form the St\"ackel matrix~$S$  \cite{ts08a,ts08,ts10,ts10b}. In this case separated relations~(\ref{seprel}) are af\/f\/ine equations in integrals of motion~$H_k$.

Let us  calculate the control matrices for the  Steklov--Lyapunov system and, for brevity,  introduce  three  constants
\[
\tau_k=\mbox{tr}\,\mathbf  A^k\equiv\sum_{i=1}^3 a_i^k ,\qquad k=0,1,2,
\]
and  some linear functions  on  variables $p$, $M$
\begin{gather}\label{rho-sigma}
\rho_k=\langle c, \mathbf A^k p\rangle,\qquad  \sigma_k=\langle c, \mathbf A^k M\rangle ,\qquad k=0,1,2,
\end{gather}
which are related to each other via  the   Casimir functions (\ref{Caz})  on $e^*(3)$. For instance,
\begin{gather}\label{caz-c1}
C_1=\sum_{i=1}^3p_i^2=-\dfrac{1}{2}\big(\tau_1^2-\tau_2\big)\rho_0^2+2(\tau_1\rho_1-\rho_2)\rho_0-\rho_1^2.
\end{gather}
In this notations  control matrices  associated with the Hamiltonians $H_{1,2}$ (\ref{int-stl}) and the  Poisson bivectors $P'_{1,2}$ (\ref{p2-p}), (\ref{p2-r}) look like
\begin{gather}\label{f-matsl}
 F_1=
       \begin{pmatrix}
        -2\rho_1  & 2\rho_0  \vspace{1mm}\\
         \sigma_1+\rho_2-\tau_1\rho_1  &-\sigma_0-\rho_1+\tau_1\rho_0  \\
       \end{pmatrix},
     \qquad
 F_2=
         \begin{pmatrix}
           \dfrac{\sigma_0-\rho_1}{\rho_0} -\dfrac{\tau_1}{3}& 2 \vspace{2mm}\\
          \dfrac{\sigma_1+\rho_2-\tau_1\rho_1 }{\rho_0}  & \dfrac{2\tau_1}{3}
         \end{pmatrix}.
\end{gather}
Now we can simply  calculate the desired  variables of separation using two control matrices.
Namely, let $u_{1,2}$ be eigenvalues of the  control matrix~$F_2$
\begin{gather}
B(\lambda) = \det(F_2-\lambda)=(\lambda-u_1)(\lambda-u_2)
\nonumber\\
\phantom{B(\lambda)}{} =
\lambda^2-\left(\dfrac{\sigma_0-\rho_1}{\rho_0}+\dfrac{\tau_1}{3}\right)\lambda
-\dfrac{2\tau_1^2}{9}+\dfrac{2(2\rho_1+\sigma_0)\tau_1}{3\rho_0}-\dfrac{2(\sigma_1+\rho_2)}{\rho_0} , \label{bpol}
\end{gather}
whereas the eigenvalues of  $F_1$ be  doubled  momenta $2p_{u_{1,2}}$, so that the  characteristic polynomial has the form
\begin{gather}
A(\lambda) = \det(F_1-\lambda)=(\lambda-2p_{u_1})(\lambda-2p_{u_2})\nonumber\\
\phantom{A(\lambda)}{} =\lambda^2+(\sigma_0+3\rho_1-\tau_1 \rho_0) \lambda+2\rho_1(\sigma_0+\rho_1)-2\rho_0 ( \sigma_1+\rho_2) . \label{apol}
\end{gather}
Another equivalent def\/inition of momenta $p_{u_i}$ is given by a relation
\[
p_{u_i} =\dfrac{\langle c,p \rangle}{2} u_i -\dfrac{\langle c,M\rangle}2+\dfrac{\langle c,d\times p\rangle}{3} ,\qquad i=1,2.
\]
Now we can prove the following

\begin{proposition}
On symplectic leaves of $e^*(3)$ variables $u_{1,2}$ and $p_{u_{1,2}}$ are canonical variables
\[
\{u_i,p_{u_i}\}=1,\qquad \{u_i,p_{u_i}\}'_1=2p_{u_i},\qquad \{u_i,p_{u_i}\}'_2=u_i,\qquad i=1,2.
\]
with respect to canonical Poisson bracket   \eqref{e3}.
\end{proposition}

The proof consists of the calculation of the Poisson brackets between coef\/f\/icients of characteristic polynomials $A(\lambda)$ (\ref{apol}) and $B(\lambda)$ (\ref{bpol}).

Now we have to  determine an  inverse transformation from variables $u_{1,2}$ and $p_{u_{1,2}}$ to initial variables~$p$,~$M$. Firstly, using the def\/initions (\ref{apol}),~(\ref{bpol}) and the relation~(\ref{caz-c1}),
we express  f\/ive linear functions~(\ref{rho-sigma}) via  variables of separation and Casimir functions{\samepage
\begin{gather}
\rho_0 = 2 \dfrac{p_{u_1}-p_{u_2}}{u_1-u_2} ,\qquad
 \rho_1= \dfrac{2\tau_1-3u_1}{3(u_1-u_2)} p_{u_1}-\dfrac{2\tau_1-3u_2}{3(u_1-u_2)} p_{u_2} ,\nonumber\\
\rho_2 = \dfrac{(u_1-u_2)C_1}{4(p_{u_1}-p_{u_2})}
-\left(\tau_2+\dfrac{\tau_1^2}{9}\right)\dfrac{p_{u_1}-p_{u_2}}{2(u_1-u_2)}
+\dfrac{2\tau_1}{3}\dfrac{p_{u_1}u_2-p_{u_2}u_1}{u_1-u_2}
+\dfrac{(p_{u_1}u_2-p_{u_2}u_1)^2}{4(p_{u_1}-p_{u_2})(u_1-u_2)} ,
\nonumber\\
\sigma_0 = \dfrac{u_1+2u_2}{u_1-u_2} p_{u_1}-\dfrac{2u_1+u_2}{u_1-u_2} p_{u_2} ,
\label{rs-up}\\
\sigma_1 = -\rho_2+\dfrac{\tau_1\rho_1}{3}-\dfrac{2(p_{u_1}u_1-p_{u_2}u_2)\tau_1}{3(u_1-u_2)}
+\dfrac{(p_{u_1}-p_{u_2}) u_1u_2}{u_1-u_2} .\nonumber
\end{gather}
 It is easy to see  that  $\rho_k$ and $\sigma_k$ are symmetric functions in $u_{1,2}$, $p_{u_{1,2}}$.}

Secondly, we determine the  initial variables $p$, $M$ as functions on $\rho_{k}$ and $\sigma_{k}$:
\begin{gather}
p_i = c_i \bigl(a_ja_k\rho_0-(a_j+a_k)\rho_1+\rho_2\bigr) ,\qquad i=1,2,3, \qquad (i,j,k)=(1,2,3),\nonumber\\
\label{inv-trans}
M_i = \dfrac{c_i}{\rho_0}\bigl(C_2-\sigma_1(a_i\rho_0-\rho_1)
+\sigma_0\bigl(\rho_2-\tau_1\rho_1+a_i(a_j+a_k)\rho_0\bigr)\bigr) .
\end{gather}
Here $c_i $ are entries of the vector $c$ (\ref{ab-v}),  $C_{1,2}$ are the  Casimir functions   on $e^*(3)$  (\ref{Caz}) and $(i,j,k)$ means  the permutation of $ (1, 2,3)$.

Matrices $F_{1,2}$ (\ref{f-matsl}) in canonical variables of separation look like
\[
F_1=S \begin{pmatrix} 2p_1&0 \\ 0&2p_2 \end{pmatrix} S^{-1} ,\qquad
F_2=S \begin{pmatrix} q_1&0 \\ 0&q_2\end{pmatrix} S^{-1} ,
\]
where the St\"ackel matrix $S$ is equal to
\begin{gather}\label{st-mat}
S=    \begin{pmatrix}
      \dfrac{6}{2\tau_1-3u_1} & \dfrac{6}{2\tau_1-3u_2} \vspace{2mm}\\
      1 & 1
    \end{pmatrix}.
\end{gather}
The notion of the St\"ackel matrix $S$ allows us to easily get the separated relations (\ref{seprel}) and prove that canonical variables $u$, $p_u$ are the variables of separation for the Steklov--Lyapunov system.
\begin{proposition}
In the Steklov--Lyapunov case the canonical variables $u_{1,2}$  \eqref{bpol} and $p_{u_{1,2}}$~\eqref{apol} satisfy  the following  separated relations
\begin{gather}\label{newsep-sl}
\Phi(u,p_u)=\left(\dfrac{u}{2}-\dfrac{\tau_1}{3}\right)H_1+H_2+\varphi_3(u) p_u^2
+\phi_3(u)=0 ,\!\qquad u=u_{1,2},\!\qquad p_u=p_{u_{1,2}} ,\!\!
\end{gather}
where cubic polynomials $\varphi_3(u)$ and $\phi_3(u)$  are equal to
\begin{gather}
\varphi_3(u)=\left(
\dfrac{u^3}{2}+\dfrac{u}{3}\big(\tau_1^2-3\tau_2\big)-\dfrac{4}{27}(\tau_1-3a_1)(\tau_1-3a_2)(\tau_1-3a_3)
\right),\nonumber\\
\phi_3(u)=\frac{C_1u^3}{2}+C_2u^2-\left(\frac{C_2\tau_1+C_1\tau_1^2}{3}-\frac{C_1\tau_2}{2}\right) u
+C_2\left(\frac{7\tau_1^2}{9}-\tau_2\right)\nonumber\\
\phantom{\phi_3(u)=}{}
+C_1\left(\frac{\tau_1^3}{27}+\frac{2\tau_1\tau_2}{3}-\frac{2\tau_3}{3}\right).\label{phi-pol}
\end{gather}
\end{proposition}

The proof consists of  substituting  integrals of motion $H_{1,2}$ (\ref{int-stl}) in terms of variables of separation (\ref{inv-trans}) into the separated relations (\ref{newsep-sl}).

So, in the Steklov--Lyapunov case equations of motion are linearized on Jacobian of  the genus two
hyperelliptic  curve def\/ined by the equation $\Phi(u,p_u)=0$ (\ref{newsep-sl}) and the system of the Abel--Jacobi equations has the standard form
\begin{gather}
\int^{u_1}_{\infty} \dfrac{du}{p(u)\varphi_3(u)}+\int^{u_2}_{\infty}\dfrac{du}{p(u)\varphi_3(u)} = \beta_1t+\gamma_1 ,\nonumber\\
\label{h-quad}
\int^{u_1}_{\infty} \dfrac{u du}{p(u)\varphi_3(u)}+\int^{u_2}_{\infty}\dfrac{u du}{p(u)\varphi_3(u)} = \beta_2t+\gamma_2 .
\end{gather}
Here  $p(u)$ means  the function $p_u$ on $u$ obtained from  the equation (\ref{newsep-sl}), $\beta_{1,2}$ are   certain constants depending only on the choice of the Hamiltonian ($H_1$ or $H_2$)  and  $\gamma_{1,2}$ are  two constants. Solving these equations  with respect to $u_{1,2}(t,\gamma_{1,2})$  and substituting these solutions into the expressions (\ref{rs-up}) and (\ref{inv-trans}) we f\/inally get  the initial variables~$p$, $M$ as functions on time variable~$t$ and six  constants $H_{1,2}$, $C_{1,2}$ and~$\gamma_{1,2}$.

\begin{remark}
In order to give an explicit theta-functions solution, one can apply the standard machinery  of the Weierstrass root functions describing inversion of the hyperelliptic quadratures~(\ref{h-quad}),  completely similar to solution of Jacobi's geodesic problem or  Neumann's particular case of the Clebsch system~\cite{bt09,fed09,w94}.
\end{remark}

Using shift  of coordinates
\[
u_i\to \dfrac{2}{3}(a_1+a_2+a_3-3u_i)
\]
we can rewrite  separated relations~(\ref{newsep-sl}) in the following  form
\begin{gather}\label{sep-22}
p_u^2=\dfrac{C_1u^3+(C_1\,\mbox{tr}\,\mathbf A-C_2)u^2+4\tilde{H}_1u+4\tilde{H}_2}{\det(u-\mathbf A)} ,
\end{gather}
where
\[
\tilde{H}_1=-H_1-C_1\,\mbox{tr}\left(\mathbf A^2+4\mathbf A^\vee\right)+2C_2\,\mbox{tr}\,\mathbf A,
\qquad \tilde{H_2}=H_2+2C_1\det\mathbf A- 2C_2\,\mbox{tr}\,\mathbf A^\vee .
\]
Separated relations (\ref{sep-22}) have the same form as the  K\"otter separated relations (\ref{sepeq-K}). However, we have to point out that if $p$ and $M$ are real variable, then~$u_{1,2}$ are complex functions in contrast with the real K\"otter variables  $v_{1,2}$. In terms of $u_{1,2}$ and $p_{u_{1,2}}$ symmetric functions on the K\"otter variables~$v_{1,2}$ look like
\[
v_1+v_2=\dfrac{P(u_1,u_2,p_{u_1},p_{u_2})}{Q(u_1,u_2,p_{u_1},p_{u_2})} ,\qquad
v_1v_2=\dfrac{R(u_1,u_2,p_{u_1},p_{u_2})}{T(u_1,u_2,p_{u_1},p_{u_2})} .
\]
Here $P$, $Q$, $R$, $T$ are the sixth-order polynomials in momenta, $Q$, $T$ are the sixth-order polynomials in coordinates,
whereas $P$ and $R$ are the seventh- and eighth-order polynomials in coordinates. We did not f\/ind a foreseeable expressions for these polynomials or their combinations. In any case variables $u$, $p_u$ and $v$, $p_v$ are related by non-trivial canonical transformation~(\ref{gen-trans}).

\begin{remark}
The Steklov--Lyapunov system on $e^*(3)$ coincides with the Steklov system on $so^*(4)$ after some linear change of phase variables~\cite{bob}.  It is a twisted Poisson map, which permutes f\/irst and second Lie--Poisson brackets on  $e^*(3)$ and $so^*(4)$~\cite{ts04b}.

We suppose that variables $u_{1,2}$ and $p_{u_{1,2}}$ coincide with the complex variables of separation for the Steklov system on $so(4)$  introduced in \cite{van00,kuz00} up to this change of variables and transformation of the canonical momenta associated with  permutation of the Poisson brackets.  We  thank one of the referees for the reference on  these papers. Algebro-geometric relations of this complex  coordinates $u_{1,2}$ with the real K\"otter coordinates $v_{1,2}$  is discussed in~\cite{fed09}.
\end{remark}

\subsection{The Rubanovsky system}

Let us  consider a nontrivial integrable generalization of the Steklov--Lyapunov system discovered by Rubanovsky \cite{rub}
 \begin{gather}\label{int-rub}
\hat{H}_1=H_1+2\langle b,p\rangle,\qquad \hat{H}_2=H_2+\langle b, (\mbox{tr}\,\mathbf A-\mathbf A) p\rangle
-\langle b,M\rangle ,
 \end{gather}
 where $H_{1,2}$ are given by (\ref{int-stl}) and $b=(b_1,b_2,b_3)$ is a constant vector.
 This deformation  describes the motion of a gyrostat in an ideal f\/luid under the action of the Archimedes torque, which arises when the barycenter of the gyrostat does not coincide with its volume center. The problem of separation of variables for the Rubanovsky systems was unsolved up until now.

The Rubanovsky integrals of motion are non-homogeneous second-order polynomials and, therefore,  it is natural  to solve the equations (\ref{bi-inv}) and (\ref{comp-eq}) in the space of non-homogeneous second-order polynomial bivectors and in the similar space of rational bivectors.
\begin{proposition}
Integrals of motion $\hat{H}_{1,2}(p,M)$ \eqref{int-rub} are in bi-involution  \eqref{bi-inv} with respect to  the  Poisson brackets associated with the  polynomial Poisson bivector
\begin{gather}\label{p1-rub}
\hat{P}'_1=P'_1+\langle b,c \rangle
                                     \begin{pmatrix}
                                       0 & 0 & 0 & 0 & 0 & 0 \\
                                       0 & 0 & 0 & 0 & 0 & 0 \\
                                       0 & 0 & 0 & 0 & 0 & 0 \\
                                       0 & 0 & 0 & 0 & p_3 & -p_2 \\
                                       0 & 0 & 0 & -p_3 & 0 & p_1 \\
                                       0 & 0 & 0 & p_2 & -p_1 & 0
                                     \end{pmatrix}
\end{gather}
and the rational Poisson bivector
\begin{gather}\label{p2-rub}
\hat{P}'_2=\dfrac{1}{\langle c,p\rangle} \hat{P}'_1+\dfrac{1}{\langle c,p\rangle} \left(\langle c,M\rangle+\dfrac{\alpha(1+\alpha^2\langle c,d\rangle)}{2\alpha^2\langle c,d\rangle} \langle c\times p,d\rangle\right) P ,
\end{gather}
where $P'_1$ is the Poisson bivector for the  Steklov--Lyapunov system \eqref{p2-p} in which  $\alpha=1$  and  vectors $c$ and $d$ are given by~\eqref{ab-v}.
\end{proposition}

 The proof is a straightforward verif\/ication of the equations  (\ref{bi-inv}) and (\ref{comp-eq}).

\begin{remark}
If $c=(c_1,c_2,\sqrt{-c_1^2-c_2^2})$ is an arbitrary vector, $d=(c_1\alpha^{-2},0,0)$ and $\alpha\to \infty$, then the Poisson bivectors $\hat{P}'_{1,2}$ yield bi-Hamiltonian structures on $e^*(3)$ associated with  the Lagrange top~\cite{ts08a}.
\end{remark}

Following the same line as previously, let us calculate the control matrices  associated with the Hamiltonians $\hat{H}_{1,2}$ (\ref{int-rub}) and the  Poisson bivectors $\hat{P}'_{1,2}$ (\ref{p1-rub}), (\ref{p2-rub})
\[
\hat{F}_1=F_1-\dfrac{\langle b,c \rangle}{2}
                                              \begin{pmatrix}
                                                0 & 0 \\
                                                1 & 0
                                              \end{pmatrix},\qquad
  \hat{F}_2=F_2-\dfrac{\langle b,c \rangle}{2\rho_0}
                                              \begin{pmatrix}
                                                0 & 0 \\
                                                1 & 0
                                              \end{pmatrix}.
\]
Eigenvalues of these matrices are new canonical variables of separation $\hat{u}_{1,2}$ and $\hat{p}_{u_{1,2}}$,
 so that
\[
\hat{F}_1=S \begin{pmatrix}  2\hat{p}_1&0 \\ 0&2\hat{p}_2\end{pmatrix} S^{-1} ,\qquad
\hat{F}_2=S \begin{pmatrix} \hat{u}_1&0 \\ 0&\hat{u}_2\end{pmatrix} S^{-1} ,
\]
where $S$ is the same St\"ackel matrix~(\ref{st-mat}). These  variables are simply  related with the previous one
\begin{gather*}
\hat{u}_{1,2}=\dfrac{u_1+u_2}{2}\pm\dfrac{1}{2}\sqrt{(u_1-u_2)^2-\dfrac{4\langle b,c \rangle}{\rho_0}} ,
\\
\hat{p}_{u_{1,2}}=\dfrac{p_{u_1}+p_{u_2}}{2}\pm\dfrac{1}{2}\sqrt{(p_{u_1}-p_{u_2})^2-\langle b,c \rangle\rho_0} .
\end{gather*}
Initial variables $p$, $M$ are the same  functions on $\hat{\rho}_{k}$ and $\hat{\sigma}_{k}$
\begin{gather*}
p_i = c_i \bigl(a_ja_k\hat{\rho}_0-(a_j+a_k)\hat{\rho}_1+\hat{\rho}_2\bigr) ,\qquad i=1,2,3, \qquad (i,j,k)=(1,2,3), \\
M_i = \dfrac{c_i}{\hat{\rho}_0}\bigl(C_2-\hat{\sigma}_1(a_i\hat{\rho}_0-\hat{\rho}_1)
+\hat{\sigma}_0\bigl(\hat{\rho}_2-\tau_1\hat{\rho}_1+a_i(a_j+a_k)\hat{\rho}_0\bigr)\bigr),
\end{gather*}
where four functions $\hat{\rho}_k(\hat{u},\hat{p}_u)=\rho_k(u,p)$ and $ \hat{\sigma}_0(\hat{u},\hat{p}_u)=\sigma_0(\hat{u},\hat{p}_u)$ is given by (\ref{rs-up}):
 \begin{gather*}
\hat{\rho}_0=2 \frac{\hat{p}_{u_1}-\hat{p}_{u_2}}{\hat{u}_1-\hat{u}_2},\qquad
\hat{\rho}_1= \frac{2\tau_1-3\hat{u}_1}{3(\hat{u}_1-\hat{u}_2)}\hat{p}_{u_1}-\frac{2\tau_1-3\hat{u}_2}{3(\hat{u}_1-\hat{u}_2)} \hat{p}_{u_2},
\\
\hat{\sigma}_0=\frac{\hat{u}_1+2\hat{u}_2}{\hat{u}_1-\hat{u}_2}\hat{p}_{u_1}-\frac{2\hat{u}_1+\hat{u}_2}{\hat{u}_1-\hat{u}_2}\hat{p}_{u_2},
\\
\hat{\rho}_2=\frac{(\hat{u}_1-\hat{u}_2)C_1}{4(\hat{p}_{u_1}-\hat{p}_{u_2})}
-\left(\tau_2+\frac{\tau_1^2}{9}\right)\frac{\hat{p}_{u_1}-\hat{p}_{u_2}}{2(\hat{u}_1-\hat{u}_2)}
+\frac{2\tau_1}{3}\frac{\hat{p}_{u_1}\hat{u}_2-\hat{p}_{u_2}\hat{u}_1}{\hat{u}_1-\hat{u}_2}
+\frac{(\hat{p}_{u_1}\hat{u}_2-\hat{p}_{u_2}\hat{u}_1)^2}{4(\hat{p}_{u_1}-\hat{p}_{u_2})(\hat{u}_1-\hat{u}_2)}
\end{gather*}
 and one function is dif\/ferent
\[
\hat{\sigma}_1=\dfrac{\langle b,c \rangle}{2}-\hat{\rho}_2+\dfrac{\tau_1\hat{\rho}_1}{3}-\dfrac{2(\hat{p}_{u_1}\hat{u}_1-\hat{p}_{u_2}\hat{u}_2)\tau_1}{3(\hat{u}_1-\hat{u}_2)}
+\dfrac{(\hat{p}_{u_1}-\hat{p}_{u_2}) \hat{u}_1\hat{u}_2}{\hat{u}_1-\hat{u}_2} .
\]
This shift of $\hat{\sigma}_1$ acts only on  $M$ variables (\ref{inv-trans}).

\begin{proposition}
In the Rubanovski case the separated relations have the following form
\begin{gather}
\hat{\Phi}(\hat{u},\hat{p}_u) = \left(\dfrac{\hat{u}}{2}-\dfrac{\tau_1}{3}\right)\hat{H}_1+\hat{H}_2
+\varphi_3(\hat{u}) \hat{p}_u^2-\left(\dfrac{\langle b,c \rangle}{2} \hat{u}^2+\dfrac{\langle b,\mathbf D c\rangle}{3} \hat{u}+\dfrac{2\langle b,\mathbf D^{\vee} c\rangle}{9}\right) \hat{p}_u\label{sep-rub}\\
\phantom{\hat{\Phi}(\hat{u},\hat{p}_u)=}{} + \phi_3(\hat{u})-\dfrac{\langle b,c \rangle}{2}\left(\dfrac{\hat{u}}{4}+\dfrac{\tau_1}{2}\right)
+\dfrac{\langle b,c \rangle\langle b,\mathbf A c \rangle}{2}=0 , \qquad \hat{u}=\hat{u}_{1,2},\qquad \hat{p}_u=\hat{p}_{u_{1,2}}.\nonumber
\end{gather}
Here  cubic polynomials $\varphi_3$ and $\phi_3$ are given by \eqref{phi-pol} and $\mathbf D=\mathrm{tr}\,\mathbf A-3\mathbf A$.
\end{proposition}

The proof consists of  substituting  integrals of motion $\hat{H}_{1,2}$  in terms of variables of separa\-tion~$\hat{u}$,~$\hat{p}_u$ into the separated relations~(\ref{sep-rub}).

 As above, we can prove that the  equations of motion are linearized on the Jacobian variety of the  genus two
hyperelliptic  curve def\/ined by~(\ref{sep-rub}). For the brevity, here we omit the explicit expressions for a base of  holomorphic dif\/ferentials and the corresponding Abel--Jacobi equations, which may be  easily obtained using modern computer algebra software.

Calculation of the separating variables and the corresponding algebraic curve for the Ru\-ba\-novski gyrostat is a new result, which allows us to make conclusion an applicability of bi-Hamiltonian methods to study f\/inite-dimensional  integrable Hamiltonian  systems.

\subsection*{Acknowledgements}

The author is grateful to the referees for a number of helpful suggestions that resulted in improvement of the article.

\pdfbookmark[1]{References}{ref}
 \LastPageEnding

\end{document}